\documentclass[12pt]{article}
\pdfoutput=1
\usepackage{jcappub} 

\usepackage{amsmath}
\usepackage{amsfonts}
\usepackage{amssymb}
\usepackage{graphicx}
\usepackage{amsmath,amssymb,amsbsy,amstext, amsthm,xcolor}

\usepackage[english]{babel}
\usepackage[utf8]{inputenc}
\RequirePackage{color}
\usepackage{colortbl}

\title{\boldmath On the Fast Random Sampling and Other Properties of the Three Point Correlation Function in Galaxy Surveys}

\author[a]{Fidel Sosa Nu\~nez}
\author[a]{and Gustavo Niz}

\affiliation[a]{Departamento de F\'isica, Universidad de Guanajuato, Loma del Bosque No.  103 Col.~Lomas del Campestre, C.P 37150 Le\'on, Guanajuato, M\'exico}

\abstract{In the forthcoming large volume galaxy surveys higher order statistics will provide complementary information to the usual two point statistics. Low variance estimators of the Three Point Correlation Function (3CPF) of discrete data count triangle configurations with vertices mixing data and random catalogues. Large density random catalogues are used to reduce the shot noise, which leads to a computational cost of one or two orders of magnitude more than the pure data histogram. 
In this paper, we explore time reductions of the isotropic 3PCF random sampling terms in periodic boxes without using random catalogues. In the first approach, based on Hamilton's construction of his famous two point estimator, we use an ad-hoc two point correlation term, while for the second procedure we construct the operators from a geometrical viewpoint, using two sides and their opening angle to describe the 3PCF triangle configurations. We map the last result to the three triangle side basis either numerically or analytically, and show that the latter approach performs best when applied to synthetic data. Moreover, we elaborate on going beyond periodic boxes, discuss other low variance n-point estimators and present useful 3PCF visualization schemes. }

\begin{document}
\maketitle
\flushbottom

\section{Introduction and Methodology}

A detailed understanding of how matter distributes in the Universe is a key to untangle some of the greatest puzzles of modern cosmology, such as the nature of dark energy, dark matter or the physics from the Early Universe.  The two point correlation function (2PCF) or its Fourier-space counterpart, the power spectrum, has proven very successful to reconstruct this matter distribution. However, as new experiments push up the amount of cosmological data with an increasing precision, further useful information could be extracted from higher point statistics. Moreover, primordial non-Gaussianities, the nature of the gravitational interactions, the scale and running of the DM-galaxy biases or any other physics of the standard cosmological model and its extensions would strongly benefit from these higher order statistics, in particular using the next level in the hierarchy: the three point correlation function (3PCF), or its Fourier counterpart, the bispectrum. 
The 3PCF in galaxy surveys dates back to Peebles and Groth \cite{Peebles,Groth} and since then numerous studies and theoretical improvements have been developed (see for example the reviews of \cite{Bhuvnesh} and \cite{Bernardeau:2001qr}). The interest on the higher statistics has largely increased in the last two decades, as can be appreciated by the numerous studies on the SDSS data or other experiments (see for example \cite{Kayo:2004fd, Marin:2013bbb, McBride:2010zn, Pan:2005ym, Nichol:2006mg, McBride:2010zp, Marin:2010iv, Wang:2004km,2011ApJ...726...13M, Kulkarni:2007qu, Jing:1998qs, 2005MNRAS.364..620G}), or some of the more recent studies \cite{Slepian:2015hca,Slepian:2016kfz, Slepian:2016nfb, Moresco:2016opr, deCarvalho:2020ftb, Tie:2019tpi}.

Different estimators of the underlying true 3PCF have been studied, such as the Szapudi and Szalay (SS) \cite{SS} or Jing and B\"orner (JB) \cite{JB}, but insufficient work has been done on exploring further possibilities as in the two point case. In particular, the SS estimator is the 3pt equivalent of the well known minimal-variance two point estimator of Landy-Szalay (LS) \cite{LS}. Although the three point structure is richer than its two point counterpart, the SS is also expected to be a good minimal variance estimator, and hence a commonly used option. We based our findings on the SS estimator but they can be easily obtained for other 3PCF estimators.

The use of $n$-point correlation functions comes with a price: the observational noise and the computational cost are bigger as $n$ increases. New experiments, such as the stage IV galaxy surveys DESI \cite{Aghamousa:2016zmz}, LSST \cite{lsst}, Euclid \cite{euclid}, or WFIRST \cite{wfirst}, will achieve better resolution and larger volumes, increasing the signal to noise ratio for higher order statistics. For the 3PCF in particular, in-cell counts which benefit from organised distance counting using Kd-trees (see for example \cite{Moore:2000ar}) or other techniques, with the usual (two point) random sampling, are not fast enough to study the large scale 3PCF for these new volume experiments. Therefore, improvements to reduce the computational time of these na\"\i ve algorithms are needed\footnote{We will not discuss here other approaches to non-Gaussian clustering such as conditional cumulants \cite{Pan:2004ks}, percolation \cite{percolation}, minimum spanning trees \cite{trees} or Minkoswki functionals \cite{Mecke:1994ax, Kerscher:1999rw}}, such as the recent work using a multipole basis decomposition \cite{Slepian} or the efficient random counting for periodic boxes \cite{Lado}. Our independently developed work extends on the latter using two different approaches. (See a similar discussion in Fourier space which also uses the multipole decomposition \cite{Philcox:2019hdi, Philcox:2020xod}).

In the case of discrete galaxy catalogues, 3PCF estimators usually count triplets between data and random catalogues. The mixed counts take one or two vertices in the data catalogue and the remaining vertex or vertices of the triangular shape on the random sample. To minimise noise from the arbitrary random sampling, and in the same fashion as in the 2pt case, one either takes a large number of random points which increases computational triplet counting, or many random catalogues and average over them. However, these are both inefficient ways of sampling the random space. A first approach to efficiently sample the random counts is to understand how an estimator of a given n-point correlation function scales with lower point correlation functions and their uncertainties, in particular, with respect to one point statistics which do not need to be zero if the true averaged density is unknown. Hamilton \cite{Ham} used this approach to construct an estimator and quantify its uncertainty in terms of 1-pt functions. We use this methodology to generalise Hamilton's estimator to any order, reducing to the JB expression \cite{JB} for the 3pt case, and use the formalism to derive an expression for the 3PCF random sampling terms in terms of the 2PCF.

An alternative approach to fast sampling the random points is to understand how the points are geometrically distributed about the data points. For the isotropic 3PCF the picture becomes clearer in the basis where the triangle configurations are parametrised by two sides and one angle \cite{fry}. We construct analytic expressions to count the random and random-data histograms, reducing enormously the computational cost of the low variance SS estimator for the 3PCF. The paper is organised as follows: in the following section, \ref{Approach1sec}, we review Hamilton's formalism for the 2PCF and extend it to the three point case. Using this formalism we obtain the first random sampling method. In section \ref{Approaches2&3sec}, we describe the geometrical approaches to obtain analytical or semi-analytical expressions for the pure random triple histograms first, and then for the mixed data-random histograms. The latter histograms can be calculated by two different approaches, resulting in methods two and three. In the subsequent section, we apply our three methodologies to synthetic data and show which method performs best. In the last section, we include some discussions on the properties of the SS 3PCF estimator, extensions of our methods to include non periodic boundary conditions, and some remarks about our findings. Finally, we include two useful appendices where we describe Hamilton's formalism applied to a general 3PCF estimator and visualization schemes for the 3PCF.

\section{Approach 1: Minimising One Point Uncertainties} \label{Approach1sec}

\subsection{Hamilton's Formalism}

In practice, to calculate correlation functions in galaxy surveys one needs estimators that reduce to the underlying true correlation function in the infinite volume limit and when selection effects are mitigated. Consequently, for finite volume catalogues individual estimators may converge differently due to one point signals or edge effects. Let us explore this in more detail using Hamilton's formalism \cite{Ham}. Consider the physical local overdensity 
\begin{equation}\label{deltadef}
    \delta=\frac{n-\bar{n}}{\bar{n}},
\end{equation}
where $\bar{n}$ is the true mean galaxy density, together with a function $W(\mathbf{x})$, which creates a finite subsample from the underlying distribution using the product of a \emph{local} selection function $\Psi(\mathbf{x})$ with the galaxy weights $w$. If the variables are properly normalised, the number of galaxies is $N=\langle n(\textbf{x}) \Phi(\textbf{x})\rangle $, with the angular brackets denoting an average over all points $\mathbf{x}$ in space. Now, let us define the following $(n,k)$-correlation functions (with $1\leq k\leq n$) as
\begin{eqnarray}
\label{Psi}
\Psi_{k}^{(n)}\left(\mathbf{r}_{1}, \mathbf{r}_{2}, \dots, \mathbf{r}_{n}\right)
&\equiv&\frac{\left\langle W_{1}W_2 \dots W_{n} \delta_{1}\delta_2 \dots \delta_{k}\right\rangle_{_P}}{\left\langle W_{1}W_2 \cdots W_{n}\right\rangle_{_P}}\\
&\equiv&\frac{\int_{_P} d^{3} \mathbf{x}_{1} d^{3} \mathbf{x}_{2} \cdots  d^{3} \mathbf{x}_{n} W\left(\mathbf{x}_{1}\right) \cdots W\left(\mathbf{x}_{n}\right) \delta\left(\mathbf{x}_{1}\right) \cdots \delta\left(\mathbf{x}_{k}\right)}{\int_{_P} d^{3} \mathbf{x}_{1} d^{3} \mathbf{x}_{2} \cdots  d^{3} \mathbf{x}_{n}W\left(\mathbf{x}_{1}\right) \cdots W\left(\mathbf{x}_{n}\right)},\nonumber
\end{eqnarray}
where $W_1W_2\cdots W_n=w_{12\cdots n}\Phi_1\Phi_2\cdots\Phi_n$ is the weighted product of the selection functions, and the nested integrals are calculated over the region $P$, defined by the polygon with vertices at $\mathbf{r}_i$ ($i=1\dots n$) with sides given by $\mathbf{r}_{ij}$ (such that $\mathbf{r}_{12}=\mathbf{r}_1-\mathbf{r}_2$, and so on). Notice that the weights do not need to be separable (\emph{i.e.} $w_{12}\neq w_1 w_2$), however, for later purposes it would be enough to consider only the finite volume effect on the sample and to assume the galaxies are equally weighted, so that $w_{12\cdots n}=1$.

Furthermore, in the case of statistical homegeneity and isotropy the degrees of freedom reduce to $3n-6$\footnote{$3n-6$ dof hold for $n>$2, since for $n=2$ one still gets one dof.}, instead of the $3n$ vectors components $r_i$, and we only care about side magnitudes (${r}_{ij}\equiv |\mathbf{r}_{ij}|$) and enough angles (${\theta}_{i}$) between sides to uniquely defined the polygon. For example, for two-points correlations (pairs) we only have one relevant distance $r_{12}$, while for three-point functions (triplets) the minimal number to defined the triangle is either three sides ($r_{12},\ r_{23},\ r_{13}$) or two sides and one angle (e.g. $r_{12},\ r_{23}$, and the angle at the second vertex $\theta_{2}$). In the case of tetragon, four sides is not enough to fully characterised the shape and six parameters are needed. One option is to use $r_{12},\ r_{23},\ r_{34}$, the angles $\theta_2$ and $\theta_3$ (which up to here characterise a planar tetragon), and a further tilting angle between the vectors $\mathbf{r_{12}}$ and $\mathbf{r_{34}}$. Extrapolating this construction for a general polygon confined to a plane, one needs $2n-3$ variables; one possibility is to use the $n-1$ sides ($r_{12},\dots,r_{n-1\ n}$) and the $n-2$ internal angles ($\theta_{2},\dots,\theta_{n-1}$), since the rest of sides and angles will be fixed. The remaining $n-3$ variables to uniquely define the polygon in 3d can be the tilting angles $\mathbf{r_{i\ i+1}}\cdot\mathbf{r_{j\ j+1}}$ (with $i+1<j$).

There are two special limits of the $(n,k)$-correlation functions defined in (\ref{Psi}). When $(n,k)=(1,1)$, it simply reduces to the measured average density given by 
\begin{equation}
    \bar{\delta}(\mathbf{r}_1)\equiv\psi_{1}^{(1)}=\frac{\left\langle W_{1} \delta_{1}\right\rangle}{\left\langle W_{1}\right\rangle},
\end{equation}
and when $k=n$, one obtains the truth n-point correlation function, 
namely
\begin{equation}
    \hat{\xi}^{(n)}\left(\mathbf{r}_1, \dots, \mathbf{r}_n\right)\equiv\psi_{n}^{(n)}=\frac{\left\langle W_{1} \cdots W_{n} \delta_{1}\cdots \delta_{n}\right\rangle_{_P}}{\left\langle W_{1} \dots W_{n}\right\rangle_{_P}}.
\end{equation}

Most estimators of n-point correlation functions of discrete samples use n-dimensional histograms which naturally define the oriented distances of the sides that form an ``n-plet'' (2-plets=pairs, 3-plets=triplets, etc.). In this language, an n-plet histogram $X_1\cdots X_n ({r}_{1},  \dots, {r}_{n})$ is 
\begin{equation}
X_1\cdots X_n (\mathbf{r}_{1},  \dots, \mathbf{r}_{n})=\langle \epsilon_1 \dots \epsilon_n W_{1}\dots W_{n} \rangle_{_P},
\end{equation}
where $X_i$ denotes that the vertex $\mathbf{r_i}$ is either on the data ($D$) or the random ($R$) samples, integrals are over the $P$ region defined above, and $\epsilon_i$ is a function for the $i$-vertex which is equal to 1 for the random catalogue ($X_i=R$) or the density field $n_i\equiv n(\mathbf{r_i})$ for the data sample ($X_i=D$). Moreover, we add a subscript "s" when the histogram is being symmetrised over all its arguments. For example, for pairs we get the usual DD, DR and RR histograms. Moreover, we assume equal size bins for the histogram variables. To exemplify the notation, consider the 2-pt histograms 
\begin{eqnarray}
\label{2d-histograms}
DD(\mathbf{r_1},\mathbf{r_2})&=&\langle n_1n_2W_{1}W_2\rangle = \bar{n}^2\langle W_{1}W_2\rangle\Big[\xi^{(2)}(\mathbf{r_1},\mathbf{r_2})+\sum_i\Psi^{(2)}_{1}(\mathbf{r_i})+1\Big]\\
DR(\mathbf{r_1},\mathbf{r_2})&=& \langle n_1W_{1}W_2\rangle =\bar{n}\langle W_{1}W_2\rangle\Big[\Psi^{(2)}_{1}(\mathbf{r_1})+1\Big],\\
DR(\mathbf{r_1},\mathbf{r_2})_s&=& \frac{1}{2}\langle (n_1+n_2)W_{1}W_2\rangle =\bar{n}\langle W_{1}W_2\rangle\left[\frac{1}{2}\sum_i\Psi^{(2)}_{1}(\mathbf{r_i})+1\right],\\
RR(\mathbf{r_1},\mathbf{r_2})&=&\langle W_{1}W_2\rangle\ ,
\end{eqnarray}
where in the last equality of all expressions we have used the $(n,k)$-correlation function (\ref{Psi}) and delta field (\ref{deltadef}). If the random and data samples have different number of objects, one could use the normalisation factor $n_{norm}$, which scales as
\begin{equation}
 n_{norm}\equiv\frac{D}{R}=\bar{n}(1+\bar{\delta}),
\end{equation}
where by $D$ and $R$ we just mean the number of objects in each sample.
For the rest of the paper, we will often consider the same number of random points as data points to simplify the expressions.
The simplest construction of the 2PCF, $\xi^{(2)}$, is to use linearly the $DD$ histogram, and then depending on different contributions of the other histograms with the adequate $n_{norm}$ factor, one would obtain different corrections to the 2PCF from the 1pt correlation functions, which in the infinite volume limit will disappear. Table \ref{table2pcf} summarises some of the most famous 2PCF estimators and their perturbative 1pt corrections, assuming $\Psi^{(2)}_{1}$ and $\bar{\delta}$ are smaller than one. Notice that Landy-Szalay (LS) and Hamilton (H) estimators are corrected at second order in $\Psi^{(2)}_{1}$ and $\bar{\delta}$, hence their low variance, particularly on large scales.

\begin{table}
\begin{center}
\begin{tabular}{| l | l | l |}
\hline
 Name & Estimator & Departure from $\xi^{(2)}$ \\ \hline\hline
 Peebles Hauser (PH) \cite{PH} & $\displaystyle\frac{D D}{n_{norm}^2R R}-1 $ & $\displaystyle\frac{\xi^{(2)}
 +\sum_{i}\Psi^{(2)}_{1,i}-2\bar{\delta}-\bar{\delta}^2}{(1+\bar{\delta})^2}$\\ \hline 
Davis-Peebles (DP) \cite{DP} & $\displaystyle\frac{D D}{n_{norm} D R_s}-1$ & $\displaystyle \frac{\xi^{(2)}
+\frac{1}{2}\sum_i\Psi^{(2)}_{1,i}-\bar{\delta}-\frac{1}{2}\bar{\delta}\sum_i\Psi^{(2)}_{1,i}}{(1+\bar{\delta})\left(1+\frac{1}{2}\sum_i\Psi^{(2)}_{1,i}\right)}$    \\ \hline
Hewett (He) \cite{Hewett} & $\displaystyle\frac{D D}{n_{norm}^2R R}-\frac{DR_s}{n_{norm}R R}$ & $\displaystyle \frac{\xi^{(2)}
+\frac{1}{2}\sum_i\Psi^{(2)}_{1,i}-\bar{\delta}-\frac{1}{2}\bar{\delta}\sum_i\Psi^{(2)}_{1,i}}{(1+\bar{\delta})^2}$ \\ \hline
Landy-Szalay (LS) \cite{LS} & $\displaystyle\frac{D D}{n_{norm}^2RR}-2\frac{DR_s}{n_{norm}RR}+1$ & $\displaystyle \frac{\xi^{(2)}
-\bar{\delta}\sum_i\Psi^{(2)}_{1,i}+\bar{\delta}^2}{(1+\bar{\delta})^2}$ \\ \hline
Hamilton (H) \cite{Ham} & $\displaystyle\frac{D D\  R R}{(D R_s)^{2}}-1$ & $\displaystyle \frac{
\xi^{(2)}
-\frac{1}{4}\left[\sum_i\Psi^{(2)}_{1,i}\right]^2}{\left[1+\frac{1}{2}\sum_i\Psi^{(2)}_{1,i}\right]^2}$ \\ \hline
\end{tabular}
\caption{\label{table2pcf} Popular two point estimators assuming the same data and random data points, and their departure, in terms of one point operators, from the truth 2PCF $\xi^{(2)}$. We drop all vertex labels and define $\Psi^{(2)}_{1,i}\equiv\Psi^{(2)}_1(\mathbf{r_i})$. }
\end{center}
\end{table}

\subsubsection{Three and higher order estimators}

A three point correlation function would need the $DDD$ histogram together with a combination of the other three possible options $DDR$, $DRR$ and $RRR$, whose structure could be tunned to obtain a low variance estimator such as Landy-Szalay or Hamilton expressions for the two point case.  In the Hamilton's notation introduced earlier, the scaling of these three-point histograms with lower point correlators is 
\begin{equation}
    \begin{aligned} DDD(\mathbf{r_1},\mathbf{r_2},\mathbf{r_3}) &=\bar{n}^{3}\left\langle W_{1}W_2W_3\right\rangle\left[\xi^{(3)}(\mathbf{r_1},\mathbf{r_2},\mathbf{r_3})+
    \sum_{i<j}\Psi_2^{(3)}(\mathbf{r_i},\mathbf{r_j})+
    \sum_i\Psi^{(3)}_{1}(\mathbf{r_i})+1\right] \\ 
    DDR(\mathbf{r_1},\mathbf{r_2},\mathbf{r_3}) &= \bar{n}^{2}\left\langle W_{1}W_2W_3\right\rangle\left[\Psi_{2}^{(3)}(\mathbf{r_1},\mathbf{r_2}) +\Psi_{1}^{(3)}(\mathbf{r_1})+\Psi_{1}^{(3)}(\mathbf{r_2})+1\right] \\ 
    D R R(\mathbf{r_1},\mathbf{r_2},\mathbf{r_3}) &=\bar{n}\left\langle W_{1}W_2W_3\right\rangle\left[\Psi_{1}^{(3)}(\mathbf{r_1})+1\right] \\ 
    R R R(\mathbf{r_1},\mathbf{r_2},\mathbf{r_3}) &=\left\langle W_{1}W_2W_3\right\rangle \end{aligned}.
\end{equation}
The symmetrised histograms $DRR_s$ and $DDR_s$ can be easily obtained from the previous expressions. A general estimator may contain non-trivial functions of the symmetrised histograms but, as it is usually done, we would only consider a linear combination of products up to a given order. We refer the reader to Appendix A for details on the expressions, but bare in mind that there are 18 possible terms assuming up to third order in the numerator. Five of those 18 parameters are easily chosen to avoid corrections on the lowest orders, leaving a family of 2 effective parameters (that are linear functions of the remaining 13). It is important to stress that the simplest $DDD/RRR-1$ operator would have a non-vanishing two-point correlation, $\Psi^{(3)}_2$, in the infinite sampling limit, or equivalently when $\bar{\delta},\ \Psi_1^{(3)}\rightarrow 0$. Of the two parameter family of low variance estimators, the most popular choices are the Szapudi-Szalay (SS) \ref{SS} and Jing-B\"orner (JB) \cite{JB} constructions, given by the following combination of triplet histograms
\begin{eqnarray}
    \label{SS}
     \xi_{S S}^{(3)} &=&\frac{D D D-3 n_{norm}D D R_s+3 n_{norm}^2D R R_s-n_{norm}^3R R R}{n_{norm}^3R R R}, \\
    \label{JB}
    \xi_{J B}^{(3)}&=&\frac{D D D[R R R]^{2}}{(D R R_s)^{3}}-3 \frac{D D R_s \ R R R}{(D R R_s)^{2}}+2,
\end{eqnarray}
where we have dropped the vertex label to shorten the expressions. These are low biased estimators, whose corrections are given by 
\begin{eqnarray} \label{SScorrection}
\hat{\xi}_{ S S}^{(3)}&=&\left(1+\bar{\delta}\right)^{-3}\left[\xi^{(3)}-\bar{\delta} \sum_{i< j} \Psi^{(3)}_{2,ij}+\bar{\delta}^{2} \sum_{i}\Psi^{(3)}_{1,i}-\bar{\delta}^{3}\right]\\ \label{JBcorrection}
\hat{\xi}_{ JB}^{(3)}&=&\left(1+\frac{1}{3}\displaystyle\sum_i\Psi^{(3)}_{1,i}\right)^{-3}\left[\xi^{(3)}-\frac{2 }{27}\sum_i\Psi_{1,i}^{(3)}-\frac{1}{3}\displaystyle\sum_i\Psi_{1,i}^{(3)}\sum_{i<j}\Psi_{2,ij}^{(3)}\right]
\end{eqnarray}
where $\Psi^{(3)}_{1,i}\equiv \Psi^{(3)}_1(\mathbf{r_i})$ and $\Psi^{(3)}_{1,ij}\equiv \Psi^{(3)}_2(\mathbf{r_{ij}})$. This methodology may be generalised to higher orders, leading to the following low bias estimator for the n-point correlation function (nPCF)
\begin{eqnarray}
\label{nHam}
\xi _ { HAM } ^ { ( n ) } = \sum _ { k = 0 } ^ { n } ( - 1 ) ^ { n - k }\binom{n}{n-k} \frac { D ^ { ( k ) } R ^ { ( n - k ) } \left( r _ { 1 } , \cdots , r _ { n } \right)_s \left[ R ^ { ( n ) } \left( r _ { 1 } , \cdots , r _ { n } \right) \right] ^ { k - 1 } } { \left[ D R ^ { ( n - 1 ) } \left( r _ { 1 } , \cdots , r _ { n } \right)_s \right] ^ { k } }
\end{eqnarray}
where $D^{k}R^{n-k}(r_1,\cdots r_n)_s$, $k=0,1,\cdots,n$ are the histograms with $k$ data vertices and $n-k$ random vertices. This expression reduces for $n=2$ and $n=3$ to the Hamilton (see \ref{table2pcf}) and (\ref{JB}) expressions respectively.
This is to be contrasted with the n-point Szapudy-Szalay (Landy-Szalay for 2pts) estimator \cite{SS}, given by $\xi _ { LS} ^ { ( n ) } =(N/R)^n$, where $N\equiv D-n_{norm}R$. Using the formalism explained here, one can show that the leading corrections to both estimator generalisations are never linear in the one-point functions $\bar{\delta}$, $\Psi^{(3)}_{1}$. The SS estimator tries to push the $\Psi^{(3)}_{1}$ term to higher orders whereas the JB structure does it for the $\bar{\delta}$. Although, there are other 3PCF estimator options which could push these one point functions to order higher than linear, none can disappear the $\bar{\delta}$ corrections (as JB) and push the $\Psi^{(3)}_{1}$ to higher order (as SS) at the same time, as shown in Appendix A. Moreover, as one considers higher order correlators the choices increase and one could use Hamilton's formalism to construct other low bias estimators.

\subsection{Random sampling using 2pt statistics}

After introducing the formalism developed by Hamilton, and a couple of popular three-point and higher point estimators, we proceed to show how one can sample the random-data histograms of the three point correlation function by only using two point estimators. 

Our starting point is the Szapudy-Szalay estimator (\ref{SS}), given its low variance and fast convergence. Towards the end of this work (Section \ref{SSVsJBsec}), we will justify this choice in more detail, but for the moment let us take this as a working assumption. In order to sample the random catalogues with the data ones to form the $DDR_s$ or $DRR_s$ histograms we need to count triplets with at least one leg in the random sample. However, to decrease the random's noise contribution one usually takes a larger number of randoms (usually around 50 times the number of data points), or many random catalogues of similar size to the data field an average over them. In both approaches the scaling with the number of triplets is larger that those of the DDD histograms. At this point one may wonder if two point expressions can more efficiently sample the randoms. 

A first approach based in the formalism we have introduced previously holds for uncorrelated weights only (and our choice of local selection functions). In this case, $w_{1...n}=\Pi_i w_i$, the correlation $\Psi^{(n)}_k$ reduces to $\Psi^{(k)}_k$, because the integrals $\int W_i$ ($i=k+1,\dots,n$) in the numerator and denominator of (\ref{Psi}) cancel. Under this assumption the correction to the Szapudy-Szalay estimator (\ref{SScorrection}) simplifies to
\begin{equation}\label{SScorrection2}
\hat{\xi}_{ S S}^{(3)}=\frac{1}{(1+\bar{\delta})^3}\left[\xi^{(3)}-\bar{\delta}\left(\sum_{i>j} \xi^{(2)}(\mathbf{r}_{ij})-\bar{\delta} \sum_{i}\Psi^{(1)}_{1}(\mathbf{r}_{i})+\bar{\delta}^{2}\right)\right].
\end{equation}
To efficiently sample the random catalogues, we would like an estimator which only calculates the $DDD/RRR$ piece (of the SS estimator (\ref{SS})) and uses a two point estimator to approximate all further terms. In other words, we are looking for a 2PCF estimator X to build up the 3PCF Szapudy-Szalay in the following way
\begin{equation}
    \hat{\xi}_X^{(3)}\left(\mathbf{r}_{1}, \mathbf{r}_{2}, \mathbf{r}_{3}\right)=\frac{D D D}{R R R}\left((\mathbf{r}_{1}, \mathbf{r}_{2}, \mathbf{r}_{3}\right)-\sum_{i< j} \hat{\xi}^{(2)}_X\left(\mathbf{r}_{ij}\right)-1 ,
\end{equation}
such that it gives the same corrections as in (\ref{SScorrection2}). By inspecting 2PCF estimators (see table \ref{table2pcf}), one could try to find the estimator that leads to the same one point expansion, in Hamilton's language, as the term above in the round brackets. One should consider there are three 2PCF contributions, one for each triangle side, and the $(1+\bar{\delta})^{-3}$ factor when looking for the correct 2PCF expression. In turns out to be the Hewett estimator \cite{Hewett}. In summary, our first method to simplify the random counting of the Szapudy-Szalay 3PCF estimator\footnote{If, instead, one chooses the JB estimator (\ref{JB}), Hamilton's 2PCF estimator has the same one point corrections, resulting in the equivalent expression to (\ref{method1}) given by
$$
\hat{\xi}_{JB}^{(3)}\left(r_{12}, r_{13}, r_{23}\right)=\frac{D D D (RRR)^2}{(D R R_s)^3}\left(r_{12}, r_{13}, r_{23}\right)-\sum_{i<j} \hat{\xi}_{H AM}^{(2)}\left(r_{i j}\right)-1.
$$
} is 
\begin{equation}\label{method1}
    \hat{\xi}_{SS}^{(3)}\left(r_{12}, r_{13}, r_{23}\right)=\frac{D D D}{n_{norm}^3 R R R}\left(r_{12}, r_{13}, r_{23}\right)-\sum_{i< j} \hat{\xi}^{(2)}_{He}\left(r_{ij}\right)-1 .
\end{equation}

\section{Approaches 2 and 3: Geometrical Schemes}\label{Approaches2&3sec}

In this section, we try to focus on periodic boundary conditions and isotropic correlation functions, hence we only care about the triangle sides. The basic idea is to reduce the random counting with analytic expressions, which together with the $DDD$ histogram, lead to the an accurate approximation of the Szapudi-Szalay 3PCF estimator. First, we explore the $RRR$ histogram, then generalise it to the $DRR$ case, and conclude with the non-trivial case of the $DDR$ counting. By the end, we will come back to the possibility of non-periodic boxes and statistical anisotropies. 

\subsection{Analytical Expression for the RRR estimator}
\label{RRR_a}

To calculate the 2PCF in a box of volume $V$ with periodic boundary conditions, it is well known that the $RR$ piece can be obtained from an analytical expression (see, for example, the documentation of the popular code CUTE \cite{CUTE}). Let us review the construction argument of this analytic expression because it will be useful for the three point case. Imagine we take a fixed pivot point in a random catalogue with $N$ points, and draw all possible pairs from it. The pairs are equally distributed along spherical shells around the pivot point, and the pair number at each shell is proportional to the density ($N/V$, with $N$ the number of points in the box) times the spherical shell volume. Moving from one pivot to all of them will imply an extra factor of $N$. This construction gets us to the histogram on an infinitesimal thick shell given by
\begin{equation}
\label{dRR}
    dRR(r)=N\left(\frac{N}{V}\right)\left(4\pi r^2\right)dr,
\end{equation}
which one needs to integrate over $r$ bins to get the final result, namely
\begin{equation}
\label{RR_T}
    RR(r)=\int_{r-\Delta r}^{r+\Delta r}dRR= \frac{N^2}{V}v(r, \Delta r),
\end{equation}
where $v(r,\Delta r)$ is the spherical shell volume given by
\begin{equation}\label{pot}
    v(r)=\frac{4 \pi}{3}\left[(r+\Delta r / 2)^{3}-(r-\Delta r / 2)^{3}\right].
\end{equation}
The random counting converges to the previous formulae in the infinite data limit ($N_R\rightarrow\infty$). Inspired by this construction, it is straight forward to derive an analytical expression for the $RRR$ histogram, which can be easily understood in the $\{r_1,r_2,\mu=\cos\theta\}$ basis for the triplets (\cite{fry}), instead of using the three triangle's sides.

As before, consider a fix pivot random point, that we call $\mathbf{r}_1$, and from there draw all possible triangles with vertices $\mathbf{r}_2$ and $\mathbf{r}_3$ in the R catalogue. With our choice of basis, the resulting isotropic histograms of triplets will be functions of the following binned variables: two distances $r_{12}\equiv|\mathbf{r}_{12}|$ and $r_{13}\equiv|\mathbf{r}_{13}|$ and the opening angle around the pivot point $\mu_1=\cos(\theta_1)=\mathbf{r}_{12}\cdot\mathbf{r}_{13}$. Because a random catalogue does not have a preferred orientation, $\mu_1$ is a random variable, hence the histogram only depends on its binning size, $\Delta\mu_1$. In other words, the monopole of a multipole decomposition in $\theta$ contains all the $RRR$ information. Notice that using the three side triangle as our variables does not lead to the same argument because the closing triangle side, $r_{23}$, is a function of the other sides, as we will discuss in what follows. As a result of the angle independence, the $RRR$ histogram only depends on the radial distributions, $N_{piv}(r)$, of the $\mathbf{r}_2$ and $\mathbf{r}_3$ points times the number of pivot points ($N$) and the width of the angle bin ($\Delta\mu_1$). Under this construction, the $RRR$ histogram gives 
\begin{eqnarray}
\label{RRRT}
RRR(r_{12},r_{13},\mu_1)
&=& -\frac{N}{2}
N_{piv}(\mathbf{r}_{12}) 
N_{piv}(\mathbf{r}_{13})
\Delta \mu_1 \\ \nonumber 
&=& 1
-\frac{N^3}{2V^2}
v(r_{12})
v(r_{13}) 
\Delta \mu_1.
\end{eqnarray}
The factor of 2 and the minus sign are due to the fact that the Cosine function is decreasing as is arguments grows and the size of the interval is $2=\cos(0)-\cos(\pi)$.
One may use the the number of bins, $n_{\mu_1}$, instead of $\Delta {\mu_1}$ by substituting $\Delta{\mu_1}$ with $2/n_{\mu_1}$. 

\begin{figure}
\begin{center}
\includegraphics[width=6.0cm]{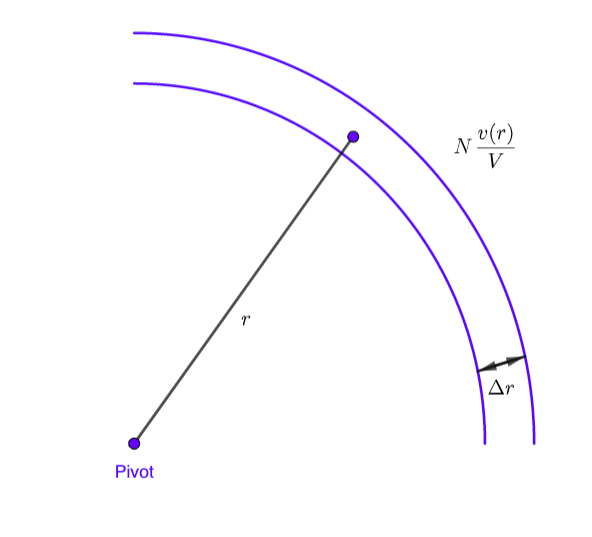}
\includegraphics[width=9.0cm]{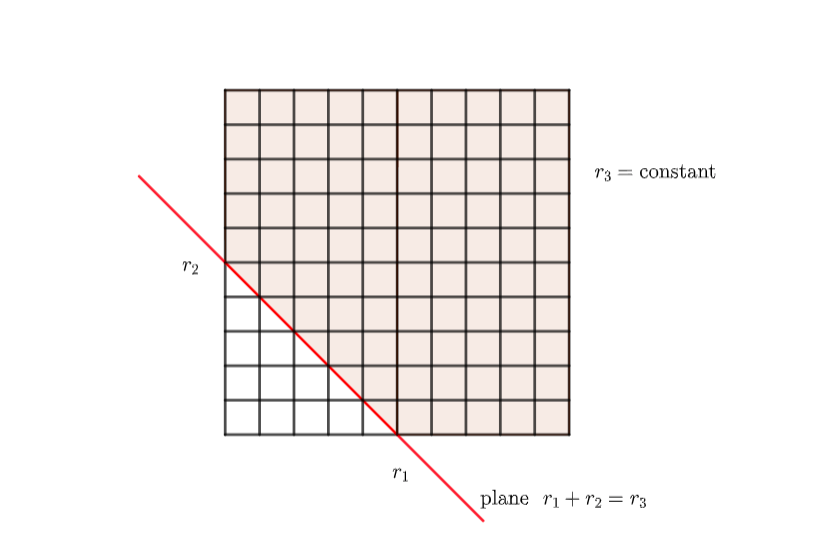}
\end{center}
\caption{{\em a) Left: 2d slice of a spherically thin  shell around a pivot point of radius $r$ and thickness $\Delta r$. The expected density of random points is $Nv(r)/V$, with $v(r)$ defined in \ref{pot}. Right: grid slice for $r_3$ fixed with forbidden region by the triangle inequality (\ref{ineq}) in white. Most errors in our fast random sampling methods arise from points in the forbidden region that fit into bins where the triangle equality (red line).}}
\label{fig1}
\end{figure}

To make further progress we need to transform this analytic RRR expression to the $r_{12}$, $r_{13}$ and $r_{23}$ basis. A simple approach is to take the infinitesimal bin limit (take a linear order limit on the ``Deltas'' in (\ref{RRRT}) and assume they are of infinitesimal size) and write the $\mu_1$ in terms of $r_{ij}$ using the law of Cosines, to substitute it back into (\ref{RRRT}). By following these steps, we obtain the infinitesimal histogram 
\begin{eqnarray}
    d R R R\left(r_{12}, r_{13}, r_{23}\right)=\frac{8 \pi^{2} N^{3}}{V^{2}} r_1r_{2} r_{3} d r_{1} d r_{2} d r_{3}
\end{eqnarray}
Then, we just integrate this result over the variable's bins, assuming the same bin size ($\Delta r$) for each side, leading to
\begin{eqnarray}\label{RRRsides}
    RRR(r_{12}, r_{13}, r_{23})&=&\int_{r_1-\Delta r/2}^{r_1+\Delta r/2}\int_{r_2-\Delta r/2}^{r_2+\Delta r/2}\int_{r_3-\Delta r/2}^{r_3+\Delta r/2}dRRR(r_{12}, r_{13}, r_{23})\nonumber\\
    &=&\frac{8 \pi^{2} N^{3}}{V^{2}} r_{12} r_{13} r_{23}[\Delta r]^{3}.
\end{eqnarray}
These integrals should be calculated over $r$-regions where the triangle inequality,
\begin{equation}\label{ineq}
r_{12}+r_{13}\geq r_{23},
\end{equation}
is satisfied. A note of caution is that a coarse binning would introduce an important error in the result. Actually, this is only relevant where the equality sign holds in the previous inequality (\ref{ineq}); see for example Figure \ref{fig1}. Geometrically, this corresponds to collinear points where the area of the triangle vanishes. To avoid this error, we can include the triangular equality in the integration limits of (\ref{RRRsides}), or consider a refined binning in those regions where the triangular equality is satisfied. We will exemplify this error and how to mitigate it in the following section, where our methods will be compared using numerical simulations.

\subsection{Analytical Expressions for the DRR and DDR crossed terms}
\label{DRR_DDR}

In this subsection, we present two different methodologies to estimate the factors $DDR_s$ and $DRR_s$, which together with the previous analytic formula for $RRR$ and the $DDD$ histogram, will be needed to complete the 3PCF Szapudi-Szalay estimator (\ref{SS}). As a reminder, these factors must be included in any 3PCF estimator to remove the bare 2PCF contribution included in the $DDD/RRR$ term.

\begin{figure}
\[\rotatebox{0}{\includegraphics[width=16.0cm,height=4.4cm]{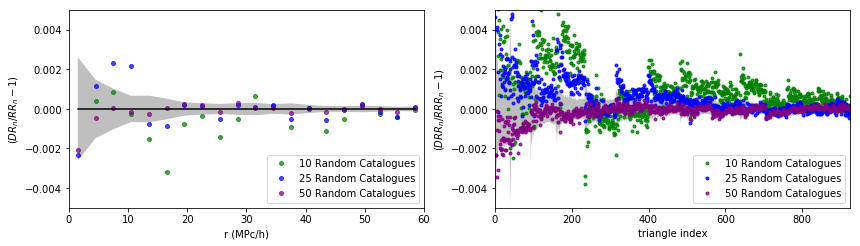}}
\]\caption{{\em As the number of random point increase for periodic boxes, the one-leg data n-point histograms ($DR_s$ in the left plot and $DRR_s$ in the right plot) converge to the purely random n-point histograms ($n_{norm}RR$ and $n_{norm}RRR$ respectively). We use the data points from the section \ref{mockcomparison} simulation, three sets of $10,$ $25$ and $50$ random catalogues and estimate the dispersion from repeating the 50 random catalogue calculation twenty five times. The random samples have the same points as the data ($n_{norm}= 1$).}}
\label{DRtoRRfig}
\end{figure}

Using periodic boxes, as well as local weights and the selection function, enormously constraints the statistical properties of some histograms in n-point correlation functions. As it is well known for the 2PCF, the $DR$ tends to $n_{norm}RR$ once the number of random points is sufficiently large (this can be thought as a consequence of achieved translation invariance in this limit). For the same reason, all n-point histograms with only one leg of the polygon in the data catalogue and the rest on the random samples ($DR\dots R$) reduce to the purely random n-point histogram ($n_{norm}R\dots R$). Therefore, for our 3PCF estimator all $DRR_s$ terms can be approximated by the $RRR$ analytic expression (\ref{RRRsides}). As previously explained, this result can be understood in Hamilton's language as the fact that all one point contributions vanish. For the 2PCF and 3PCF cases, the details on how the one-leg data histograms converge to the pure random ones is better appreciate in Figure \ref{DRtoRRfig}. For the 2PCF this fact implies that all histograms of Table \ref{table2pcf} reduce to the Peeble-Hauser expression. In contrast, the vanishing one point correlations do not imply all 3PCF estimators are equal, instead, there are non-trivial two point pieces remaining from the $DDD$. In the SS estimator, the removal of such tow point contribution is mostly encoded in the $DDR_s$ term. Therefore, we will focus on this term for the rest of our analysis, since it is not trivial to obtain analytically. 

Following the same construction as in section \ref{RRR_a} for the $RRR$ histogram, we can develop an equivalent expression for the $DDR_s$ histogram. We start, as before, in $\{r_{12},r_{13},\mu_1\}$ basis where the intuition helps for the initial derivation. Afterwards we need to map our result to $\{r_{12},r_{13},r_{23}\}$ basis in order to obtain the final result. Consider a pivot point, $\mathbf{r}_{1}$, in the data catalogue with $N$ points and draw all possible triangles with one further vertex in the data points, $\mathbf{r}_{2}$, and another in the random sample, $\mathbf{r}_{3}$, which we also assume to contain $N$ points. As before, the angle $\mu_1$ is randomly distributed so the $DDR(r_{12},r_{13},\mu_1)$ only depends on the product of two independence pair distributions. The $\mathbf{r}_{12}$ pairs have both legs on the data points, hence the distribution cannot be approximated analytically and it proportional to $DD(r_{12})$. In contrast, the $\mathbf{r}_{13}$ pairs are randomly distributed about the pivot point and given as before by $N_{piv}(r)$. The resulting expression is 
\begin{equation}\label{DDRnosym}
    D D R\left(r_{12}, r_{13}, \mu_{1}\right)=-\frac{N}{2 V} D D\left(r_{12}\right) v\left(r_{13}\right)\Delta\mu_1,
\end{equation}
where, as before, the $-2$ is due to the range and decreasing property of the Cosine function.
Notice that this histogram is not necessarily symmetric with respect to interchanging the first two variables, but that can be easily obtain by symmetrising it. However, this result can not be the full $DDR_s$ histogram, which should contain triangles where the pivot points, $\mathbf{r}_{1}$, are placed in the random catalogue and the other vertices in the data points. In the case of a "R" pivot point, the opening angle is not a random variable anymore, and our line of thought cannot be followed. Nevertheless, our $DDR$ result (\ref{DDRnosym}) contains enough the information to build up the reaming pieces, since it can be mapped to the $(r_{12},r_{13},r_{23})$ basis and symmetrise it in those variables. In order to do so we follow two possible methodologies, which lead to different precision results.

\subsubsection{Approach 2: numerical completion of $DDR$}\label{method2sec}

The first of these methodologies uses the law of cosines numerically to compute the closing triangle side $r_{23}$ and map the $DDR$ of equation (\ref{DDRnosym}) into the $(r_{12},r_{13},r_{23})$ basis. Then we just symmetrise over all variables and obtain the final $DDR_s$ histogram. The main drawback of this approach is that uncertainty on each distance bin before the mapping may enlarge the error on the resulting $r_{23}$. To minimize this error, we can take smaller bins ($\Delta r^0 < \Delta r$ and $\Delta\mu_1^0<\Delta\mu_1$) in $r_{12}$, $r_{13}$ and $\mu_1$ initially and the re-bin our final result to the desired size. This algorithm has a complexity of order $(2\pi/\Delta\mu^0_1)( r_{max}/\Delta r^0)^2$.

\subsubsection{Approach 3: analytic completion of $DDR$}\label{method3sec}
In this method, we follow the same procedure in section (\ref{RRR_a}) to map the histogram to the $(r_{12},r_{13},r_{23})$ basis and symmetrise it in that space. For that, as we did for the $RRR$, it is easier to construct a differential for of equation (\ref{DDRnosym}), obtaining
\begin{eqnarray}
       d D D R\left(r_{12}, r_{13}, r_{23}\right)&=&-\frac{8 \pi^{2} N^{3}}{V^{2}}\left(1+\hat{\xi}^{(2)}\left(r_{12}\right)\right)r_{13}^2r_{13}^2dr_{12}dr_{13}d\mu_1\nonumber\\
       &=&\frac{8 \pi^{2} N^{3}}{V^{2}}\left[1+\hat{\xi}^{(2)}\left(r_{12}\right)\right] r_{12} r_{13} r_{23} d r_{12} d r_{13} d r_{23}\ ,
\end{eqnarray}
where we have used that $DD(r)\sim  RR(r)\left(1+\xi^{(2)}(r)\right)$ (remember all 2PCF estimators converge to the same form for periodic boxes, since $DR\rightarrow RR$), the analytic expression (\ref{RR_T}) for $RR$, taken only linear terms on each $dr_{ij}$, and the law of Cosines to obtain the last line in the three triangle sides' basis.
We integrate this differential expression over a bin of size $\Delta r$ on each triangle side, resulting in
\begin{equation}
         D D R\left(r_{12},\! r_{13},\! r_{23}\right)=R R R\left(r_{12},\! r_{13}, \! r_{23}\right)+\frac{8 \pi^{2} N^{3}}{V^{2}} r_{13} r_{23}[\Delta r]^{2}\!\! \int_{r_{12}-\frac{\Delta r}{2}}^{r_{12}+\frac{\Delta r}{2}} \! \!\! \! \! \!\! \! r_{12}  \hat{\xi}^{(2)}\left(r_{12}\right) d r_{12}
\end{equation}
To evaluate the integral, we can either interpolate the 2PCF and do the integral numerically, or approximate it with a sum over very small bins to avoid errors. We will use the later, and even for a very fine-grained binning the calculation would not add many further computational resources compared to the DDD calculation. Notice that this histogram is not symmetric yet, thus after symmetrising the expression and adding the $DDD$ piece, we obtain our final 3PCF estimator formula
\begin{equation}\label{method3}
         \xi_{SS}^{(3)}\left(r_{12},\! r_{13},\! r_{23}\right)=\frac{D D D}{n_{norm}^3 R R R}\left(r_{12},\! r_{13},\! r_{23}\right)-\sum_{i<j} \frac{1}{ r_{ij}} \overline{r_{ij}\xi^{(2)}(r_{ij})}-1,
\end{equation}
where the expression with the overbar is the average of $r\xi^{(2)}$ over the coarse-grained bin $\left[r_{ij}-\frac{\Delta r}{2}, r_{ij}+\frac{\Delta r}{2}\right]$ using the fine-grained bins. Moreover, where the triangle equality holds there is a further error and we need, hence further refinements in the bins are needed.

\section{Applying our Fast Sampling Methodologies to Mock Catalogues}\label{mockcomparison}

In this section we aim at applying the previous three methodologies to synthetic data and asses the degree of accuracy. For this purpose we use the L-PICOLA code \cite{Howlett:2015hfa} to generate a small simulation with $N=32^3$ particles and a fixed volume of $V=(250 \mathrm{MPc/h})^3$. Since our aim is not to get a highly accurate signal but to find out which of the different methods performs best, it is enough to consider this box size and density. 

In order to appreciate the difference between models we use different plotting techniques. On a first approach, we map all bins (a 3d array) into a one dimensional object by moving along one variable completely holding the other two variables fixed. We then jump to the next bin in the second variable and repeat the first variable bin-sweeping. We generalise this idea into the third variable, and call the 1d axis the \emph{triangle index}. This technique is particularly useful to evaluate how different 3PCF contributions differ from each other while taking into account errors, as shown in Fig. \ref{DRtoRRfig}. However, it does not perform well to understand particular features of the signal, since adjacent bins in the 3d bin array may not be adjacent in the triangle index, resulting in periodic jumps of the signal. A second, and popular, visualization scheme is obtained by fixing the third triangle side $r_{23}$ to a set fixed values and do 2d density plots on the remaining triangle sides $r_{12}$ and $r_{13}$. It is hard to asses differences and errors compared to the signal with this technique. An alternative approach to this last one is to use the opening angle $\mu_1$ instead of $r_{23}$ and do a multipole decomposition (based on Legendre polynomials) on that angle \cite{Szapudi:2004gg}, which captures more information but is less intuitive. We refer the reader to Appendix B for further details about 3PCF visualizations.

\begin{figure}
\[\rotatebox{0}{\includegraphics[width=14.0cm]{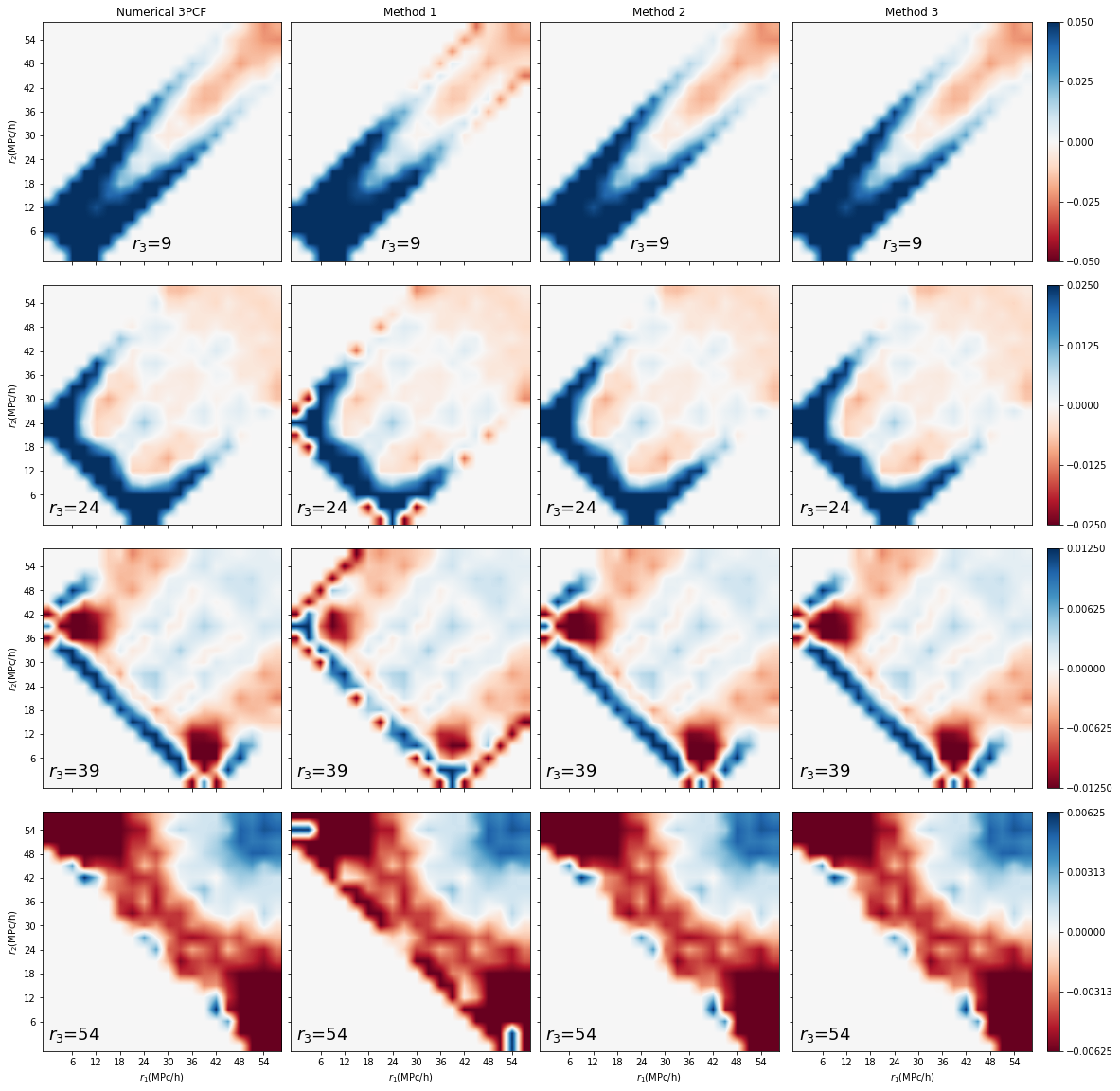}}
\]\caption{{\em 3PCF in the configurations space using the SS estimator. Each column shows 3PCF slices with $r_3$ fixed ($r_3=9,\ 24,\ 39,\ 54$). The allowed $r_1-r_2$ region is determined by the triangle inequality (\ref{ineq}). The first column corresponds to the $SS$ estimator using 100 random catalogues while the remaining columns are the three fast random sampling methods developed here. Method 1 is based on Hamilton's formalism (eq. (\ref{method1b})), while Methods 2 (section \ref{method2sec}) and 3 (section \ref{method3sec}) are the geometrical approaches. By eye, the overall agreement is good, particularly for the two geometrical methods. The largest failures are for the first method at the boundaries of the allowed region, where the triangle equality holds. Figure \ref{DDRFig} shows in more detail how each method performs, and for larger simulation boxes see the similar plot of \ref{projectionslarge}.}}
\label{projections}
\end{figure}

How do the methodologies presented before match the 3PCF Szapudy-Szalay estimator? Our results can be graphically summarised in Figure \ref{projections}. All methods do give approximate the same signal, however, there are small errors, mostly in the short scale regime, associated to the details of each method that we describe in what follows. In order to test our fast correlation function expressions in this small periodic box, we first calculate the 3PCF numerically using the SS estimator (\ref{SS}). For this, we use 1250 random catalogues with same mean density of particles as in the data ($n_{norm}=1$), and average over 25 groups of 50 random samples each to obtain the mean of different histograms of the 3PCF and their dispersion. In Fig. \ref{DRtoRRfig} we show the convergence for the DRR histograms depending on a partial number of used random catalogues, together with the magnitude of the final dispersion using the largest number of randoms catalogues (50). Moreover, we consider a maximum distance of $r_{max}= 60 \mathrm{MPc/h}$ ($24\%$ of the box side length) and bin sizes of $\Delta r=3\mathrm{MPc/h}$. In order to achieve accurate results and avoid using interpolation schemes, we take smaller bins in the geometrical approaches (methods two and three) at different triangle configurations. We discuss the details of each method in what follows, but before doing so, keep in mind that we assume an integer number of smaller bins in each $\Delta r$, to avoid interpolating between the coarse and fine grained grids.

\begin{figure}
\[\rotatebox{0}{\includegraphics[width=16.0cm,height=8cm]{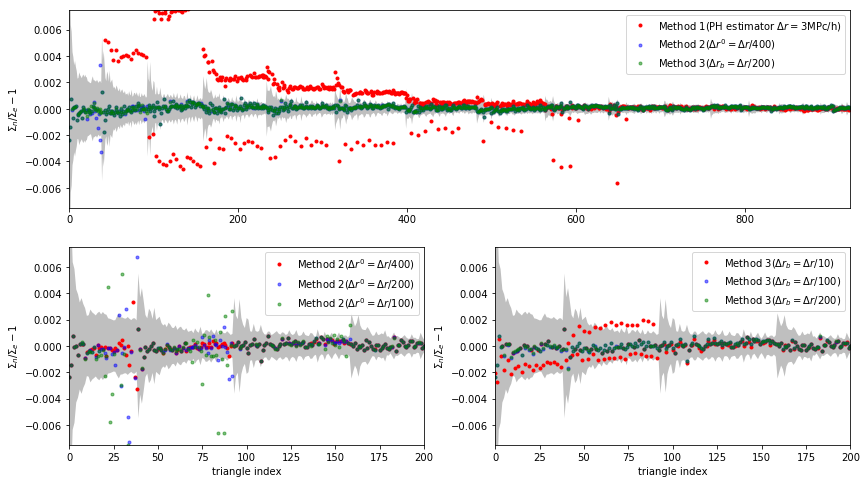}}
\]\caption{{\em The top plot shows the normalised differences between the numerical piece of the 3PCF with random-data mixing terms ($\Sigma=(3DDR-3DRR+RRR)/RRR$) and the 3 methods to fast sampling the randoms catalogues. The bottom left graph describes how the second method \ref{method2sec} converges as one increases the fine-grained binning in the two point statistics. The bottom right plot displays the convergence of third methodology \ref{method3sec} as one decreases the bin size on those r's that satisfy the triangle equality (equal sign in (\ref{ineq})). Errors are estimated using 25 independent groups of 50 random catalogues.}}
\label{DDRFig}
\end{figure}

For the first method, one can notice that the two point estimator is the same as the basic one of PH (\cite{PH}, Table \ref{table2pcf}) because, as it has been discussed before, using periodic boxes implies that the $DR$ histograms reduce to the $RR$ one. Moreover, to avoid any random sampling in the two point contribution we can calculate the $DD$ piece and use the analytic formula for $RR$, given by (\ref{RR_T}). Therefore, the resulting formula based on our previous construction (\ref{method1}) for the first method is 
\begin{equation}\label{method1b}
    \hat{\xi}_{SS,1}^{(3)}\left(r_{12}, r_{13}, r_{23}\right)=\frac{D D D}{R R R}\left(r_{12}, r_{13}, r_{23}\right)-\frac{V}{N^2}\left[\frac{DD\left(r_{12}\right)}{v(r_{12}, \Delta r)}+\frac{DD\left(r_{13}\right)}{v(r_{13}, \Delta r)}+\frac{DD\left(r_{23}\right)}{v(r_{23}, \Delta r)}\right]-1,
\end{equation}
where $v(r)$ is given by (\ref{pot}). In our study we calculate the $DD$ using the same binning $\Delta r$ as for the numerical 3PCF. This method approximates well at large scales but has larger departures at short and intermediate scales (see Figures \ref{DDRFig} and \ref{projections}). Actually, it strongly fails where the triangle equality is satisfied, as it is appreciated by eye in Figure \ref{projections}, where one sees the deviation (shown as the opposite color to the numerical 3PCF) on the boundary allowed region of each plot (for the second column corresponding to method one). Those boundaries are precisely where the triangle equality holds. If we were to work with larger and more physical mock catalogues, the deviations would be very important, as we discuss in Appendix B, particularly in Figure \ref{projectionslarge}. It is important to stress that in this method there is not any refinement of the grid used in the 2 point statistic part to improve the approximation, as opposed to the other two methodologies that we describe next. This is because we are not dealing with histograms but with a correction that is the 2PCF. 

For the second method, we map the $DDR$ histogram (\ref{DDRnosym}) from $(r_{12},r_{13},\mu_{1})$ to $(r_{12},r_{13},r_{23})$ using the law of Cosines, which by construction satisfies the triangle inequality (\ref{ineq}). However, because of this nonlinear mapping we need a refined binning in the two point pieces to diminish the associated error in the 3PCF final expression. By using three choices for the fine-grained grid given by $\Delta r^0= (1/100,\ 1/200,\ 1/400)\Delta r$, in Figure \ref{DDRFig}, we show how the methodology converges towards the 3PCF signal within its numerical error. Moreover and for simplicity, we assume the same number of bins in the angular and radial directions, hence  $\Delta\mu=\frac{2\Delta r^0}{d_{max}}$. As a result of our trials, we recommend having at least two order of magnitude difference between coarse and fine grained bin sizes, $\Delta r$ and $\Delta r^0$. To discuss about the computational cost of this method, notice that the final calculation -- without counting the $DDD$ piece -- scales as the cube of the number of fine-grained bins from the mapping (remember that the $DD$ histograms could be included as a secondary product of the $DDD$ calculation), which is smaller than any algorithm that counts triplets between random and data catalogues.

In the third and final method, we take the formula (\ref{method3}), and consider two  additional partitions or re-binnings in order to avoid precision errors of the final result. In the first refinement we take a finer grained grid to calculate the $r\xi^{(2)}$ average in expression (\ref{method3}). For this re-binning we use $\Delta r^0=\Delta r/400$ and notice that these additional bins do not increase the total calculation time by much, because they only apply to the two point statistics. The result of this approach can be seen in the top plot of Figure (\ref{ineq})), where the only points that do not fit into the expected 3PCF numerical calculations are those where the triangle equality holds (the equal sign of (\ref{ineq})). Pictorially, we can see this error arising as those triangular sections on the boundary region in Figure \ref{fig1}. In order to improve the signal on those points we perform a second refinement, but on the corresponding bins only. In Figure \ref{DDRFig} we choose this additional rebinning to be $\Delta r^{eq}=(1/10,\ 1/100,\ 1/200)\Delta r $ and see how the result converges to the full numerical 3PCF if this second re-binning is sufficiently small.

Finally, an independent work \cite{Lado} constructed a similar fast random sampling method to our geometrical approaches, but based on a different construction directly in the $\{r_{12},r_{13},r_{23}\}$ basis. Their formulae arise from the intersection of two spherical thin shells, about the centers $\mathbf{r}_1$ and $\mathbf{r}_2$, with radii $r_{13}$ and $r_{23}$ respectively. Details can be found in \cite{Lado}, but to summarise they find the formula \begin{equation}
\begin{aligned}
XXR_s\left(r_{12}, r_{13}, r_{23}\right)=4 \pi N^3 V\left\{\int_{r_{12}-\Delta r / 2}^{r_{12}+\Delta r / 2} n_{XX}\left(r_{12}^{\prime}\right) r_{12}^{\prime 2} V_{\mathrm{inter}}dr_{12}^{\prime}+permutations\right\},
\end{aligned}
\end{equation}
where $n_{norm}=1$ and each $X$ could be either $D$ or $R$. The function $V_{inter}(r_{12},r_{13},r_{23})$ is the intersection volume of the two spherically thin shells, which also depends on $\Delta r$, whose analytic expression can be found in \cite{Lado}. The density number is $n_{RR}=n_{RD}=1$, but for $n_{DD}$, one uses an interpolation of the $DD$ histogram, to get  $n_{DD}(r)=Interpolate(DD,r)/(3 N\, v(r))$, where $v(r)$ is defined by (\ref{pot}). For the $RRR$ and $DRR_s$ histograms, the authors arrived to our same expressions however, we differ in the conceptually most complicated histogram, the $DDR_s$. Moreover, because of these constructions the authors conclude that their formalism does not perform well in bins there the triangle equality holds.

\begin{figure}
\[\rotatebox{0}{\includegraphics[width=16.0cm,height=8cm]{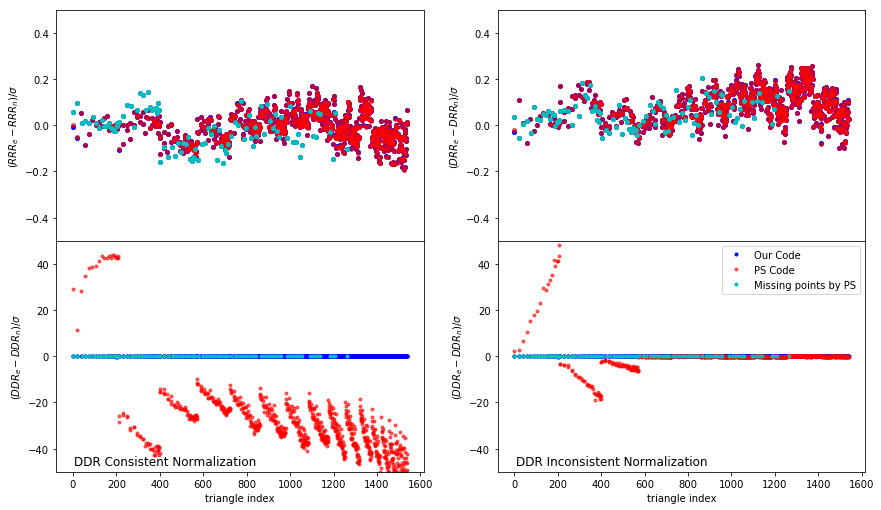}}
\]\caption{{\em Comparison between our best performing method (number 3) and the one proposed in \cite{Lado}. Differences between the numerical histograms and the fast-sampling random methodology are shown, weighted by the dispersion of the numerical 3PCF estimation using 100 random catalogues. If one chooses an appropriate normalization, the RRR and DRR histograms (top left and right respectively) are the same, except for bins where the method of \cite{Lado} fails (i.e. triangle equality region). In contrast, for the same normalization, the DDR histogram by \cite{Lado} departs on all scales (bottom left). A different normalization that adjusts the large scales, still shows deviations on shorter scales (bottom right).}}
\label{LadoVsUsFig}
\end{figure}

We use their open-source code to assess the accuracy compared to our best method: number 3. Moreover, we use the same variable as in \cite{Lado} to show the difference in performance, which is given by the simple difference of the analytic method with the numerical operator, divided by the dispersion of the latter. Results are shown in Figure \ref{LadoVsUsFig}, where we can see that our method outperforms theirs. There is a normalization factor that makes the $RRR$ and $DRR_s$ results to agree. However, for this same normalization factor their $DDR$ signal strongly disagrees with the numerical expression, hence from our result as well. If one uses a different normalization factor only for their $DDR_s$ signal, then their result agrees on large scales, but again deviates considerably on the short scale regime, apart from the collinear points where their code does not give an answer. We believe our precision lies on using a basis with one angle instead of the third triangle side, which allow us to push the analytic expression further (Eq.~(\ref{method3})) than with the other methods.

\section{Discussion}

Before stating some conclusions on our findings, we would like to discuss two important extensions to our methodologies: how the low variance 3PCF estimators converge with respect to a numerical random sampling, and how some of our results can be extended to more general finite volume surveys.

\subsection{Assessing estimators convergence with the number of random points}\label{SSVsJBsec}

\begin{figure}
\[\rotatebox{0}{\includegraphics[width=10cm]{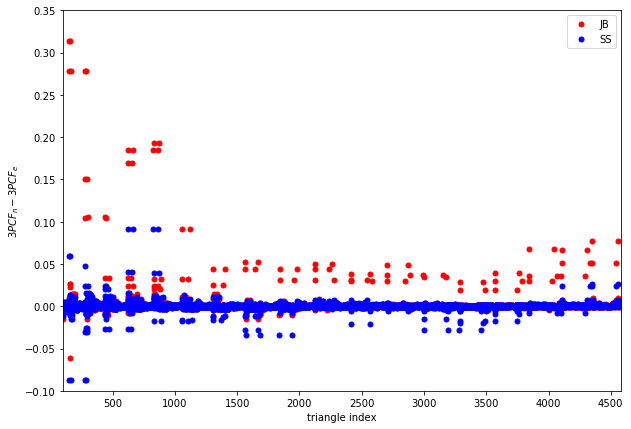}}
\]\caption{{\em Relative differences between the $3PCF_n$, estimated from the SS estimator (Eq.~(\ref{SS})) or the JB (Eq.~(\ref{JB})) estimator, and the analytic expression $3PCF_e$ of our third fast random sampling method (Eq.~(\ref{method3})). With 10 random catalogues, the SS estimator converges faster than the JB one.}}
\label{SSVsJBFig}
\end{figure}

We have chosen to work with the Szapudi-Szalay (SS) estimator instead of other choices such as the Jing-B\"orner (JB). However, after the tools and results presented here, we are in a place to justify our choice. Among all low variance estimators (see Appendix A for details), the structure of SS estimator has leading order corrections which depend on the biased overdensities $\bar{\delta}$ (see Eq.~\ref{SScorrection}), and not on the other one point function, $\Psi^{(3)}_1$, as it happens with the JB expression (Eq.~\ref{JBcorrection}). The expectation that $\bar{\delta}$ has a lower shot noise contribution from the number of random points, suggests that the SS estimator converges faster than JB, or any other estimators whose leading correction has $\Psi^{(3)}_1$ terms. Notice, however, that the SS leading correction in Hamilton's language may be obtain by other estimators (see Appendix A for a discussion on the degeneracy of these corrections). Furthermore, we could use the third analytic method for random sampling to asses the fast convergence of the SS estimator over JB counterpart. For periodic boundary conditions, since the histogram $DRR$ converges to $RRR$, the SS (\ref{SS}) and JB (\ref{JB}) reduce to the same expression, namely
\begin{equation}
    \xi^{(3)}(r_{12},r_{13},r_{23})=\frac{DDD}{RRR}-3\frac{DDR_s}{RRR}+2.
\end{equation}
However, provided a not so large number of random points, the convergence to the final result is different for the two estimators. If we assume the third methodology gives an accurate answer of the final asymptotic expression (as we have discussed in the previous section), then we can compare how the SS and JB estimators converge given the number of random points. The results are summarised in Figure \ref{SSVsJBFig}, where it is clear how the SS estimator converges  faster than the JB one. Now that we have justified the SS estimator, we could think of using our results in arbitrary volumes.

\subsection{Fast random sampling with non periodical boundary conditions}

Can we apply our calculations for the random sampling pieces of 3PCF to a non-periodic finite volume with arbitrary shape? In principle, there are a few extensions to our hypotheses that one can follow (see examples of methodologies to address the edge effects \cite{book1, ripley, Pan, SS, Slepian}). One idea corresponds to divide the arbitrary volume, $V$, into a central region, $W$, where all spheres drawn from each point in $W$ of the maximal desired length, $r_{max}$, fit entirely into the volume $V$. In this region $W$, our fast random sampling formulae would be valid as long as the weight on each galaxy remains local and there are no further weights on the random catalogues\footnote{This would not be the case in surveys such as DESI where one could use pair weighting to mitigate the assignment of optical fibers to the targets \cite{Bianchi:2018rhn}.}. In consequence, all boundary effects would come from the thick boundary region ($Q\equiv V-W$) where the signal would be obtained in the usual way using random catalogues. Under this construction, the histogram $DDD$ is calculated over the whole volume, but the $XXR$ pieces (with $X$ either $D$ or $R$) are the addition of two parts: $XXR_{Q}+XXR_W$. The pieces with sub-index $Q$ are obtained numerically by counting enclosed triangles in the volume $W$, whereas those in $W$ come from our analytical expressions, with the following details. For the $RRR_W$ histogram, one uses the analytical expression (\ref{RRRsides}), but with $r_{12}$ being within $W$ and $r_{13}$, $r_{23}$ running over the whole volume $V$. In the $DRR$ case, one gets $DRR_W=RRR_W$, as we have discussed previously. Finally, if we consider the third methodology (section \ref{method3sec}), the $DDR_W$ histogram is obtained using the expression (\ref{method3}), with the restriction that $r_{12}$ is confined to $W$ but not the other distances. The final result of these $W$-histograms is obtained after symmetrising over the three distances. This 3PCF algorithm looses speed over a periodic box, but still has the benefit of the analytic expressions in the $W$ region, which would introduce an important time reduction for large and simply connected survey volumes.

A further speed push may be achieved by using a adaptive boundary $W*$, whose size depends on the maximal scale involve in a given triangle configuration \cite{book1}. For a particular bin with $r_{12}$, $r_{13}$ and $r_{23}$ in the 3PCF, the size of $W*$ is such that all spheres of radius $Max(r_{12},r_{13},r_{23})$ about any point in $W*$ are fully contained within $V$. As in the fixed volume case, our analytic expression can be used in this regime with the restriction of $r_{12}\in W*$, and symmetrising the final results over the three distances. The triangles in the complement space ($Q*=V-W*$) are obtained using random catalogues. This algorithm becomes particularly effective when all scales of the triangle are small, leaving the full triplet random-data counting for triangles where at least one of the scales is large.\\
Although we have outlined a border correction strategy to shown how these fast sampling formulae can be applied to more general volumes, one still needs to address the inclusion of optimal and systematic weights, among other issues (such as redshift space), to really apply these techniques to more realistic galaxy surveys. However, the results presented here are useful at many stages of real data analysis, such as to validate the perturbation theory modeling of the signal and the assessment of some systematic errors.

\section{Conclusions}

The use of higher statistics in new large volume galaxy surveys, as complementary information to the two point clustering measurements, will be greatly pursued in the forthcoming years. However, efficient algorithms to overcome the calculation scaling with the number of objects of these correlation functions are needed to fully exploit their use. In this work, we focus on the three point correlation function (3PCF) and propose three methodologies to count triplets of mixed random and data points without the need of random catalogues. These methodologies directly extend to higher correlation functions by including not only the two point pieces but also the whole hierarchy of lower correlation functions up to one order less than the desired result.\\ 
In the 3PCF case, the pure data histogram $DDD$ contains the connected three point clustering (times the $RRR$ histogram), together with a non-trivial piece of two point correlations if the object weights are local. In order to single out the connected three point piece one needs to include data-random mixing histograms. Szapudi-Szalay and Jing-B\"orner choose particular combinations that ensure further subleading one point functions show up at quadratic order. These corrections become important for finite volume surveys. A first idea is to substitute these random-data mixed terms with a two point estimator that has the same one point corrections. We call this our method 1 and the two point piece turns out to be the Hewett estimator. The result has a poor convergence to the full 3PCF result, especially in the spread limit when the triangle vertices become collinear. However, this idea is useful to construct other low variance estimators (see Appendix A).\\
A different strategy for periodic boxes is to understand how random points distribute about data points. The final distribution is isotropic, hence does not depend on opening angle between triangle sides. This approach allow us to derive a pure random histogram, which is also the result when one triangle vertex is in the data catalogue for periodic boxes. The interesting histograms is when two triangle vertices are in the data catalogue, since it is this piece that contains the two point correlations. We arrive at finite analytic expressions in the two sides and one angle basis and map them to the three sides basis.

Finally, since our fast sampling prescriptions scales as the number of objects squared, it is possible to introduce these ideas into the Slepian-Eisenstein 3PCF code \cite{Slepian}, which also has a quadratic scaling, to obtain an even faster 3PCF code. 

\section*{Acknowledgement}
We acknowledge the support of DAIP-UG, Instituto Avanzado de Cosmología A.~C., and CONACyT (in particular, for FS graduate's scholarship and research grant No. 286897). We also appreciate the computer resources of DCI-UG DataLab and Atócatl-UNAM. Finally, we thank Octavio Valenzuela, Oleg Burgueño and Julio César Clemente González for technical support, Baojiu Li for let us use his simulations, and Daniel Eisenstein and Zachary Slepian for sharing their 3PCF code.

\appendix

\section{Construction of Low Variance 3PCF Estimators}

In this section we show in detail how to obtain the SS (\ref{SS}) and JB (\ref{JB}) estimators from a general 3PCF expresion using Hamilton's formalism (Section ). The most general 3PCF estimator combines terms with powers of the histograms $DDD$, $DDR_s$, $DRR_s$ and $RRR$. If we assume at most cubic order terms in the denominator and $n_{norm}=1$, the most general estimator would be given by
\begin{eqnarray}
    \label{3pcfgeneral}
    \begin{aligned}
    \xi^{(3)}_{EST}&=a_{0} \frac{D D D\ D R R^{2}}{R R R^{3}}+a_{1} \frac{D D D\ R R R^{2}}{D R R^{3}}+a_{2} \frac{D D R\ D R R^{2}}{R R R^{3}}+a_{3} \frac{D D R\ R R R^{2}}{D R R^{3}}\\
    &\quad+a_{4} \frac{D R R^{3}}{R R R^{3}}+a_{5} \frac{R R R^{3}}{D R R^{3}}+a_{6} \frac{D D D\ R R R}{D R R^{2}}+a_{7} \frac{D D D\ D R R}{R R R^{2}}+a_{8} \frac{D D R\ R R R}{D R R^{2}}\\
    &\quad+a_{9} \frac{D D R\ D R R}{R R R^{2}}+a_{10} \frac{D R R^{2}}{R R R^{2}}+a_{11} \frac{R R R^{2}}{D R R^{2}}+a_{12} \frac{D D D}{R R R}+a_{13} \frac{D D D}{D R R}\\
    &\quad+a_{14} \frac{D D R}{R R R}+a_{15} \frac{D D R}{D R R}+a_{16} \frac{D R R}{R R R}+a_{17} \frac{R R R}{D R R}-a_{18}.
    \end{aligned}
\end{eqnarray}
Different one or two point corrections to the 3PCF estimators depend on the  tuning of the coefficients $a_i$. There are three natural conditions we can initially impose: that the $\xi^{(3)}$ coefficient is one, a vanishing constant term and no isolated two point contributions. These conditions resume, respectively, in
\begin{gather}
a_{0}+a_{1}+a_{6}+a_{7}+a_{12}+a_{13}=1, \quad \quad  
\sum_{1\leq i\leq 17}a_i-a_{18}=0,\\
3 a_{0}+3 a_{1}+a_{2}+a_{3}+3 a_{6}+3 a_{7}+a_{8}+a_{9}+3 a_{12}+3 a_{13}+a_{14}+a_{15}=0.
\end{gather}
By analyzing the remaining terms, we get a general three-point estimator:
\begin{eqnarray}
\begin{aligned}
\xi^{(3)}_{EST}&=\xi^{(3)}\left[1+b_1\sum_{1\leq i\leq3}\Psi^{(3)}_{1}(r_i)-3(1+b_1)\bar{\delta}\right]\\
&\quad +b_3\sum_{1\leq i\leq3}\Psi^{(3)}_{1}(r_i)-3b_3\bar{\delta}+b_4\bar{\delta}\sum_{1\leq i\leq3}\Psi^{(3)}_{1}(r_i)\\
&\quad+\sum_{1\leq i<j\leq3}\Psi_2^{(3)}\left[b_2\sum_{1\leq i\leq3}\Psi^{(3)}_{1}(r_i)-(1-3b_2)\bar{\delta}\right]\\
&\quad+\mathcal{O}\Big[\bar{\delta}^p\big(\sum_{1\leq i\leq3}\Psi^{(3)}_{1}(r_i)\big)^q\Big]_{p+q\geq 2}\ , 
\end{aligned}
\end{eqnarray}
where 
\begin{eqnarray}
\begin{aligned}
    b_1&=\frac{1}{3}\left(2 a_{0}-3 a_{1}-2 a_{6}+a_{7}-a_{13}\right)\\
    b_2&=\frac{1}{9}\left(6 a_{0}-9 a_{1}+2 a_{2}-3 a_{3}-6 a_{6}+3 a_{7}-2 a_{8}+a_{9}-3 a_{13}-a_{15}\right)\\
    b_3&=\frac{1}{3}\left(5 a_{0}+4 a_{2}-a_{3}+3 a_{4}-3 a_{5}+a_{6}+4 a_{7}+3 a_{9}+2 a_{10}-2 a_{11}\right.\\
    &\quad\quad\quad\quad\left.+3 a_{12}+2 a_{13}+2 a_{14}+a_{15}+a_{16}-a_{17}\right)\\
    b_4&=\frac{1}{3}\left(-25 a_{0}-16 a_{2}-a_{3}-9 a_{4}-9 a_{5}-a_{6}-16 a_{7}-9 a_{9}-4 a_{10}\right.\\
    &\quad\quad\quad\quad\left.-4 a_{11}-9 a_{12}-4 a_{13}-4 a_{14}-a_{15}-a_{16}-a_{17}\right)\ .
\end{aligned}
\end{eqnarray}
The last term of the RHS converges faster to zero than the other terms in the large volume limit, hence it can be neglected, resulting in a reliable 3PCF estimator with 15 ($18-3$) free parameters. A natural further choice that lowers the variance is $b_3=b_4=0$, leaving a family of two effective parameters $b_1$ and $b_2$, whose values are degenerated among the remaining 13 free combinations of the $a_i$'s. At this level, all estimator are corrected at second order and the corrections cannot be made all zero. Therefore, further restrictions would allow us to choose among these low variance estimators for the 3PCF. There are two choices of parameters worth mentioning: 1) if only the linear term contributions in the numerator are considered ($a_k=0$ for $k\leq11$), and by imposing a leading second order correction in  $\sum\Psi_1^{(3)}$ and $\bar{\delta}$ ($b_1=b_2=0$), we obtain the Szapudy and Szalay (\ref{SS}) estimator, meanwhile 2) using all terms up to cubic order in the numerator, and demanding a leading sixth order correction in $\bar{\delta}$ ($b_1=-1,\ b_2=1/3$) results in the Jing and B\"orner (\ref{JB}) estimator. However, notice that there are other options that lead to the same quadratic corrections of JB or SS, which we do not intend to explore further in here. Furthermore, choosing to have $\Psi^{(3)}_1$ corrections instead of $\bar{\delta}$ would naturally lead to estimators that converge slower with the number of random points. We elaborate more on this in Section \ref{SSVsJBsec}.

\section{3PCF Visualization Schemes }

In the literature, we can find several visualization schemes for the 3PCF. Here, we focus on the isotropic correlation function case, which depends on three variables; thus the visualization is not a trivial task. Below we discuss a few schemes to visualize the 3PCF, its advantages and disadvantages. \\

\begin{figure}
\[\rotatebox{0}{\includegraphics[width=14.0cm]{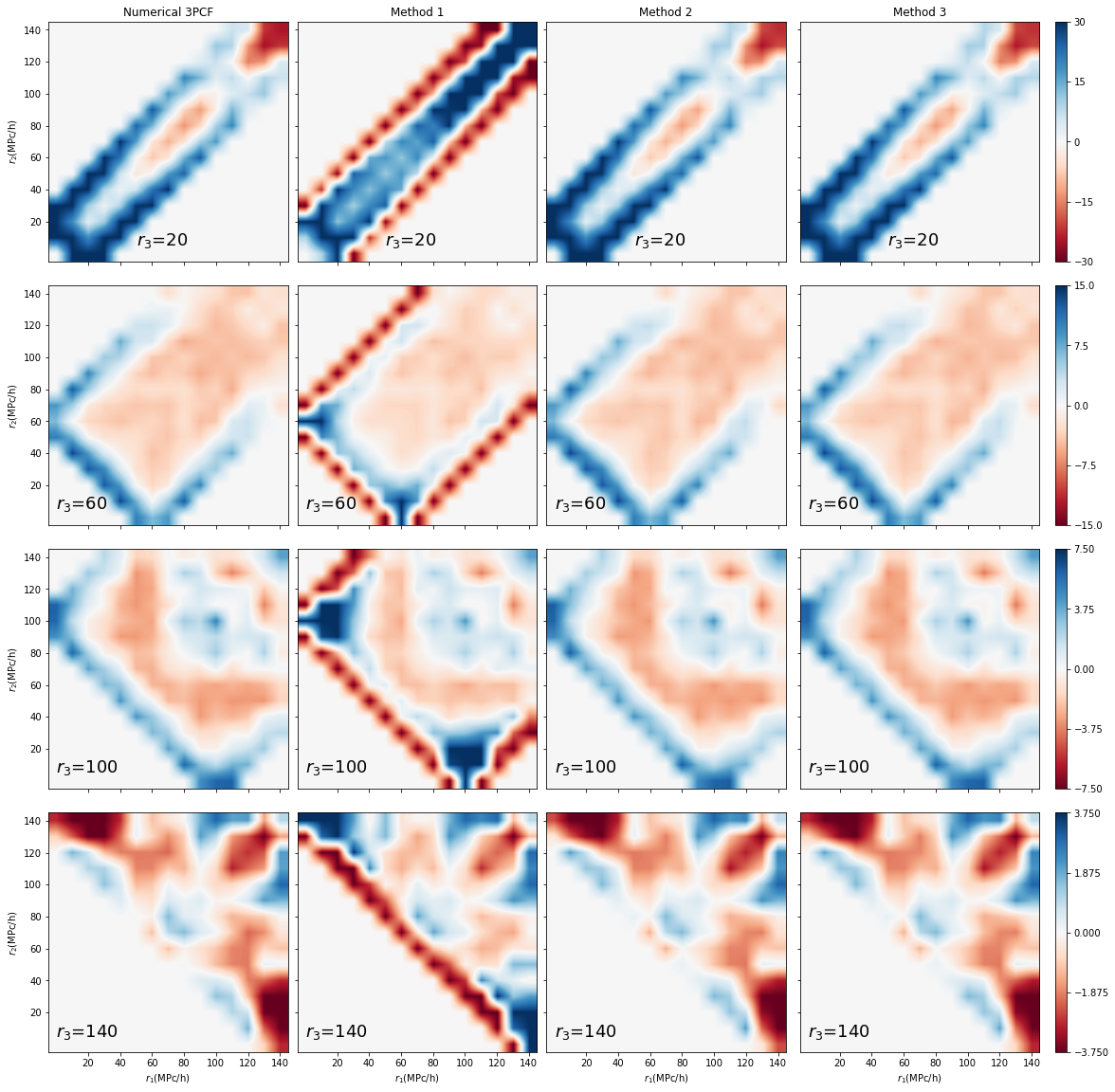}}
\]\caption{{\em Same figure as \ref{projections} but with a larger mock catalogue of halos from a larger simulation with $x^3$ particles in a box of $(1GPc)^3$. We use 50 random catalogues for the numerical 3PCF and a binning of $\Delta r=$. For the method one we do not take smaller bins, for method two we use a $\Delta r^0=$ in all bins and for method 3 a $\Delta r^0=$ where the triangle equality holds.}}
\label{projectionslarge}
\end{figure}

In configuration space (with 3 sides of the triangle), the 3PCF defines a tetrahedral allowed region. A popular choice in Fourier Space is to sketch the Bispectrum as a transparent 3d density plot (see for example \cite{tetra}). Although this scheme would allow a full 3PCF visualization, the superposition of information on the image can complicate the study of the information, specially far away from border of the allowed region where densities from different bins superpose. To avoid this visual saturation, a popular choice is to take one relation between variables (ej. $r_{13}$ or $r_{12}/r_{13}$ fixed to a constant) to obtain a 2d density plot, or even better, a 2d scatter plot when a further relationship among the variables is considered. Actually, these particular triangle configurations, if chosen before constructing the histograms, may accelerate the calculation, since not all possible triangles would be scanned by the algorithm. Geometrically, the equilateral (all sides equal), squeezed (isosceles with one side tending to zero) and spread (towards the collinear points) configurations stand out, and due to their symmetry these shapes could also pick up strong signals from perturbation theory. However, to consider special triangles decreases the available information on the 3PCF, which may be relevant for some studies, such as testing gravity \cite{c3, aviles}, where the larger difference between models may not be where the largest 3PCF signal is. Figure \ref{projectionslarge} shows a example of such visualization, where we have study the performance of our three random sampling methods in a halo mock constructed from a cosmological simulations with a larger volume and a maximum scale of 140 $Mpc/h$ using the ECOSMOG code \cite{ecosmog} with Rockstar \cite{rockstar} to find halos (particular details on the mocks can be found in \cite{c3}). The result is similar to that of the smaller box in Figure \ref{projections}. One could appreciate for this larger box how to our three analytical methods perform when at least one triangle side is the BAO scale. We have a poor random sampling due to large computational cost for our computing resources, given the large box and maximal scale. However, under these conditions one can easily see how the method based on Hamilton's formalism fails almost everywhere by eye, but specially for the boundary regions where the triangle equality holds, consistently with the discussion in section \ref{mockcomparison}.

Extending on the idea of information compression from the previous paragraph, one may think of a cosmological model, which has a certain number of parameters, and search for triangular configurations that lead to the largest variance on those parameters. To our knowledge this approach has not been fully explored in the 3PCF, but there is some studies on this direction for the bispectrum (see for example \cite{Gualdi:2017iey, Gualdi:2018pyw}).

\begin{figure}
\[\rotatebox{0}{\includegraphics[width=10.0cm]{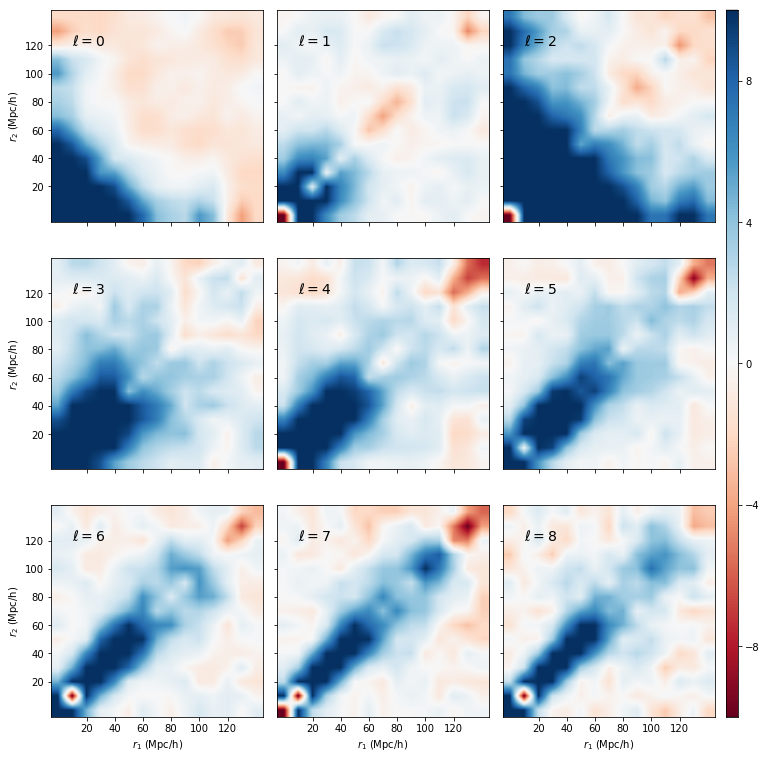}}
\]\caption{{\em Multipole coefficients $\xi_{\ell}(r_{12},r_{23})$ for several values of $\ell$, using the fast 3PCF multipole calculation code of Slepian and Eisenstein for the large mock catalogue of halos of Figure \ref{projectionslarge}. We use 50 random catalogs, and multiply the coefficients by $r_{12}\, r_{23}$ } to increase the signal on large scales.}
\label{Sle_fig}
\end{figure}

Another visualization scheme that we can find in literature is the triangle index (see for example \cite{Lado}). This scheme allows us to visualize the 3PCF as a 2d scatter plot. This is particularly relevant to compare 3PCF signals (ej. between theories, to check convergence, etc.), and specially useful when there are errors for each bin involved. The construction is simple: one just needs to glide over the cubic matrix of the radial variables in any possible way to construct a simple 1d (longer) vector. Because of this mapping, many continuous bins may be separated at the end vector, hence the signal would show periodical jumps. For this reason, it is not a useful visualization to appreciate shapes in the signal, such as the BAO structure. Examples of such visualization are in Figures \ref{DRtoRRfig}, \ref{DDRFig}, \ref{LadoVsUsFig} and \ref{SSVsJBFig}.
The third scheme we want to exemplify was first discuss in \cite{Szapudi:2004gg}. As noted by author, the isotropic 3PCF can be characterized by two sides lengths and the angle between them, $\xi^{(3)}(r_1,r_2,\cos\theta)$, which allows for multipole decomposition in a similar way as for the anisotropic 2PCF. In detail, the 3PCF decomposes into
\begin{eqnarray}
\label{multipolo_f}
\xi^{(3)}\left(r_{1}, r_{2},\cos\theta\right)=\sum_{\ell} \xi_{l}\left(r_{1}, r_{2}\right) P_{l}\left(\cos\theta\right)\ ,
\end{eqnarray}
where the coefficients $\xi_{\ell}(r_{1}, r_{2})\ (\ell=0,1,2,\cdots)$ can be easily seen as 2d density plots. These scheme can also be used to speed up the 3PCF calculation \cite{Slepian}, since one may use the relation between spherical harmonics and Legendre polynomials to an $\mathcal{O}(N^2)$ scaling of the code, as opposed to the $\mathcal{O}(N^3)$ performance of na\"{\i}ve algorithms. In Figure \ref{Sle_fig} we show how the result of calculating the multipole decomposition 3PCF signal directly using the code by Daniel J. Eisenstein and Zachary Slepian \footnote{A public version of the code \cite{Slepian} can be found in \cite{nbodykit}, $https://nbodykit.readthedocs.io/$.}.

\addcontentsline{toc}{section}{References}
\bibliographystyle{utphys}
\bibliography{main3pcf.bib}

\end{document}